\documentclass[a4paper,aps,prd,10pt,preprintnumbers,showpacs,twocolumn,superscriptaddress,nofootinbib,amsmath,amssymb]{revtex4-1}
\usepackage[dvips]{graphics}
\usepackage[utf8]{inputenc}
\usepackage[T1]{fontenc}
\usepackage{cmap}

\def\imo{i}

\def\im#1{Im(#1)}
\def\Order#1{{\cal O}\left(#1\right)}
\def\Lagrangian{{\cal L}}

\def\D{{\cal D}}
\def\P{\Psi}
\def\PD{\bar{\Psi}}
\def\gg{\gamma}
\def\ge{\hat{\gamma}}

\def\B{\tilde{B}}

\def\F{\tilde{F}}
\def\G{\tilde{G}}

\renewcommand{\section}[1]{\textbf{#1.}}\renewcommand{\appendix}{\textbf{Appendix:} }\renewenvironment{acknowledgments}[1]{\textbf{Acknowledgements.} #1}{}

\newenvironment{compactfigure}{\begin{figure*}}{\end{figure*}}

\newcommand{\arxivonly}[1]{#1}

\begin{document}
\title{Traversable wormholes in General Relativity}

\author{R. A. Konoplya}\email{roman.konoplya@gmail.com}
\affiliation{Research Centre for Theoretical Physics and Astrophysics, Institute of Physics, Silesian University in Opava, Bezručovo náměstí 13, CZ-74601 Opava, Czech Republic}

\author{A. Zhidenko}\email{olexandr.zhydenko@ufabc.edu.br}
\affiliation{Research Centre for Theoretical Physics and Astrophysics, Institute of Physics, Silesian University in Opava, Bezručovo náměstí 13, CZ-74601 Opava, Czech Republic}
\affiliation{Centro de Matemática, Computação e Cognição (CMCC), Universidade Federal do ABC (UFABC),\\ Rua Abolição, CEP: 09210-180, Santo André, SP, Brazil}

\begin{abstract}
In \cite{Blazquez-Salcedo:2020czn} asymptotically flat traversable wormhole solutions were obtained in the Einstein-Dirac-Maxwell theory without using phantom matter.
The normalizable numerical solutions found therein require a peculiar behavior at the throat: the mirror symmetry relatively the throat leads to the nonsmoothness of gravitational and matter fields. In particular, one must postulate changing of the sign of the fermionic charge density at the throat, requiring coexistence of particle and antiparticles without annihilation and posing a membrane of matter at the throat with specific properties. Apparently this kind of configuration could not exist in nature. We show that there are wormhole solutions, which are asymmetric relative the throat and endowed by smooth gravitational and matter fields, thereby being free from all the above problems. This indicates that such wormhole configurations could also be supported in a realistic scenario.
\end{abstract}
\maketitle

\section{Introduction}
Wormholes are hypothetical objects connecting disparate points of spacetime or even different universes \cite{Visser:1995cc}. Wormholes have never been observed and even their existence and formation scenarios are highly disputable questions. Nevertheless, the chance to have a traversable wormhole or construct it in a laboratory in the distant future pays off the efforts of theoreticians, attracting a lot of attention recent years. Existence of humanly traversable wormholes requires gravitational repulsion, which usually could be supported by matter with negative kinetic terms, restraining the throat from shrinking. Examples of wormholes without adding such phantom matter come at the price of modifications of the gravitational theory \cite{Maldacena:2020sxe,McFadden:2004ni,Bronnikov:2002rn,Bronnikov:2003gx,Bronnikov:2019sbx,Kanti:2011yv,Kanti:2011jz,Richarte:2007zz,Tomikawa:2014wxa}. Frequently, wormholes in such theories are unstable against linear perturbations \cite{Cuyubamba:2018jdl,Gonzalez:2008wd}. Miniature self-supported wormholes could possibly exist due to vacuum polarization in their vicinity \cite{Harko:2013yb}. Cylindrical wormhole solutions found in \cite{Bronnikov:2018uje} are noncompact and glued with the asymptotically flat spacetime.

Therefore, the crucial question is whether asymptotically flat traversable wormholes could exist as compact objects within the Einstein gravity without adding phantom matter. In this case normal matter fields must anyway violate the null energy conditions \cite{Morris:1988tu,Bolokhov:2021fil}. Until the recent work \cite{Blazquez-Salcedo:2020czn}, no solutions of Einstein equations with usual matter fields were known to provide existence of such wormholes. Wormhole solutions in Einstein gravity with added Maxwell and two Dirac fields with the usual coupling between them were found in \cite{Blazquez-Salcedo:2020czn}. Two kinds of wormhole solutions were represented there: The first one is an analytical solution, describing symmetric relative the throat wormhole supported by massless and neutral fermions, which, themselves, are nonsymmetric relative the throat. However, the fermions do not decay at infinity and are, therefore, non-normalizable\footnote{Note that traversable wormholes in the four-dimensional anti-de Sitter spacetime can be supported by massless fermions, which are localized near the throat \cite{Maldacena:2018gjk}.}. The other, normalizable, solution was obtained numerically and corresponds to symmetric configuration of both the metric tensor and matter fields. The solution was obtained by integrating the field equations between the throat and infinity, and requiring the mirror symmetry, what led to other ``exotic'' properties (see \cite{Danielson:2021aor} for details):
\begin{enumerate}
\item The throat becomes a special point where a massive shell of some matter must be posed.
\item The infinitely thin shell separates the fermion particles and antiparticles which, therefore, must meet at the throat without annihilation.
\item The metric tensor and matter fields are not continuously differentiable at the throat (although the metric and Riemann tensors are continuous).
\end{enumerate}
Thus, the consistent quantum description of such classical configuration is evidently impossible.

In this context, we are interested to know whether traversable wormholes can exist in a more realistic situation, i.~e., without the above exotic factors, such as the mass shell on the throat or coexistence of particles and antiparticles without annihilation.
Here we show that there are nonsymmetric, relative the throat, continuously differentiable solutions that describe asymptotically flat, traversable wormholes supported by normalizable and smooth matter fields. Thus, our solutions are free from all of the above disadvantages of \cite{Blazquez-Salcedo:2020czn}.

\section{Basic equations}
We consider the action \cite{Blazquez-Salcedo:2020czn}
\begin{equation}\label{action}
S=\frac{1}{4\pi}\int \sqrt{-g}\left(\frac{1}{4}R+\Lagrangian_M+\Lagrangian_1+\Lagrangian_2\right)d^4x,
\end{equation}
where the Lagrangians for the Maxwell and two Dirac fields with mass $\mu$ are defined as follows:
\begin{eqnarray}\label{Maxwell0}
\Lagrangian_M&=&-\frac{1}{4}F_{\mu\nu}F^{\mu\nu}, \qquad F_{\mu\nu}\equiv\partial_{\mu}A_{\nu}-\partial_{\nu}A_{\mu};\\
\label{Dirac1}
\Lagrangian_1&=&\frac{\imo}{2}\PD_1\gg^{\mu}\D_{\mu}\P_1-\frac{\imo}{2}(\PD_1\gg^{\mu}\D_{\mu}\P_1)^*-\imo\mu\PD_1\P_1;\\
\label{Dirac2}
\Lagrangian_2&=&\frac{\imo}{2}\PD_2\gg^{\mu}\D_{\mu}\P_2-\frac{\imo}{2}(\PD_2\gg^{\mu}\D_{\mu}\P_2)^*-\imo\mu\PD_2\P_2.
\end{eqnarray}
A spherically symmetric configuration is given by the following line element and four-potential:
\begin{subequations}\label{ansatz}
\begin{eqnarray}\nonumber
  &&ds^2=-N(x)^2dt^2 + \frac{r'(x)^2}{B(x)^2}dx^2 + r(x)^2(d\theta^2 + \sin^2\theta d\varphi^2),\\
  &&A_{\mu}dx^{\mu}=V(x)dt.
\end{eqnarray}
We employ the following ansatz for the spinors (cf.~\cite{Herdeiro:2017fhv}):
\begin{eqnarray}\label{spinoransatz1}
\Psi_1 &=& e^{-\imo\omega t+\imo\varphi/2}\left(\begin{array}{r}
\phi(x)\cos\frac{\theta}{2} \Bigr.\\
\imo\kappa\phi^*(x)\sin\frac{\theta}{2} \Bigr.\\
-\imo\phi^*(x)\cos\frac{\theta}{2} \Bigr.\\
-\kappa\phi(x)\sin\frac{\theta}{2} \Bigr.
\end{array}\right),\\ \label{spinoransatz2}
\Psi_2 &=& e^{-\imo\omega t-\imo\varphi/2}\left(\begin{array}{r}
\imo\phi(x)\sin\frac{\theta}{2} \Bigr.\\
\kappa\phi^*(x)\cos\frac{\theta}{2} \Bigr.\\
\phi^*(x)\sin\frac{\theta}{2} \Bigr.\\
\imo\kappa\phi(x)\cos\frac{\theta}{2} \Bigr.
\end{array}\right),
\end{eqnarray}
with $\kappa=\pm1$ and
\begin{equation}
\phi(x)=e^{\imo\pi/4}F(x)-e^{-\imo\pi/4}G(x),
\end{equation}
where $F(x)$ and $G(x)$ are real.
\end{subequations}
The nonzero component of the current is
\begin{equation}\label{currentexplicit}
j^0=\frac{4|\phi(x)|^2}{N(x)}=\frac{4(F(x)^2+G(x))^2}{N(x)}.
\end{equation}

Varying the action (\ref{action}) and substituting (\ref{ansatz}) in the field equations, we obtain a set of the ordinary differential equations for functions $N(x)$, $B(x)$, $V(x)$, $G(x)$ and $F(x)$ (see Supplemental Material for details).

\section{Wormholes with $Z_2$ symmetry}
Junction conditions at the throat lead to the choice of the opposite signs for $\kappa$ on different sides of the throat and changing the sign of one of the fermion functions, $G(x)$ or $F(x)$, at the throat, which corresponds to the transformation $\phi_+\to\pm\imo\phi_-^*$ \cite{Blazquez-Salcedo:2020czn}. Therefore, it is convenient to associate the two sides of space relative the throat with the opposite signs of $\kappa=\pm1$.

The analytic solution given in \cite{Blazquez-Salcedo:2020czn} suggests the appropriate choice of the compact coordinate,
\mbox{$x=\kappa\sqrt{1- r_0/r},$}
so that the two signs of $x$ describe the wormhole on both sides of the throat located at $x=0$ ($r=r_0$). Without loss of generality, we take $r_0=1$ and measure all the dimensional quantities in units of the wormhole radius.

In order to obtain a symmetric wormhole, we solve the field equations for $x>0$ ($\kappa=1$) and employ the above junction condition to produce a symmetric solution. One can check that the equations are automatically satisfied for $x<0$ ($\kappa=-1$) if
\begin{equation}\label{matching}
\begin{array}{rcl}
N_-(x)&=&-N_+(-x),\\
B_-(x)&=&-B_+(-x),\\
V_-(x)&=&-V_+(-x),
\end{array}
\qquad
\begin{array}{rcl}
F_-(x)&=&-F_+(-x),\\
G_-(x)&=&G_+(-x),\\
\omega_-&=&-\omega_+.
\end{array}
\end{equation}
Note that the mirror symmetry requires changing of the sign of frequency $\omega$ and lapse function $N(x)$ at the throat. The latter implies that the charge density (\ref{currentexplicit}) also changes its sign, i.~e., particles and antiparticles meet at the throat.
Since the Maxwell potential is an odd function of $x$, the electric strength $V'(x)$ is even, having the extreme value at the throat. Therefore, we require that
\begin{equation}\label{Wcond}
V''(0)=0.
\end{equation}
With the additional condition (\ref{Wcond}), for the fixed field charge $q$ and mass $\mu>0$, all the series coefficients for the functions $N(x)$, $B(x)$, $V(x)$, $F(x)$ and $G(x)$ can be calculated in terms of the following four parameters \arxivonly{(see the Appendix)}
\begin{equation}\label{initialcond}
\begin{array}{rclcrcl}
n_i &\equiv& N(0), &\qquad& b_i &\equiv& B'(0), \\
f_i &\equiv& F(0)\sqrt{r_0}, &\qquad& g_i &\equiv& G(0)\sqrt{r_0}.
\end{array}
\end{equation}
We use the series expansions to calculate values of the functions near the throat and use them as initial values for numerical integration. In order to solve the system of six first-order differential equations \arxivonly{(\ref{finaleqs})} it is sufficient to use the standard Mathematica\textregistered{} Livermore Solver with the quadruple-precision floating-point arithmetics. We have checked that, within the numerical tolerance of $10^{-6}$, the explicit Runge-Kutta method with increased floating-point precision yields the same results, including the asymptotic behavior of the functions in a vicinity of the singular points $x=\pm1$. Therefore, we are convinced that all six decimal cases in the presented numerical data are accurate.

Considering fixed $n_i$, $b_i$, and $f_i$, and varying $g_i$, we find that the fermion fields $F(x)$ and $G(x)$ diverge as
$$\lim_{x\to1}F(x)=\pm\infty, \qquad \lim_{x\to1}G(x)=\mp\infty,$$
changing the sign at certain values of $g_i$. Thus, one can use the shooting method to find the value $g_i$, such that $F(x)$ and $G(x)$ vanish as $x\to1$. The convergent solution is such that $B(1)=1$ and $N(1)=\sigma$, so that the asymptotic observer time is $\tau=\sigma t$. The values of the asymptotic mass $M$, charge $Q$, and post-Newtonian parameter $\gamma$ can be read off from the asymptotic behavior of the functions
\begin{equation}\label{asymptotic}
\begin{array}{rcl}
N_+(x)&=&\sigma\left(1-\dfrac{2M}{r_0}(1-x)+\Order{1-x}^2\right),\\
B_+(x)&=&1-\gamma\dfrac{2M}{r_0}(1-x)+\Order{1-x}^2,\\
V'_+(x)&=&\sigma\left(\dfrac{2Q}{r_0}+\Order{1-x}\right).
\end{array}
\end{equation}
It follows that variation of the parameter $n_i$ scales $\sigma$, and we fix $n_i$ in such a way that $\sigma=1$ ($t=\tau$). Following \cite{Blazquez-Salcedo:2020czn}, we choose $f_i=0$. Then $b_i$ parametrizes a family of wormholes with $Q>M$, approaching the extremally charged Reissner-Nordström black hole in the limit $b_i\to0$. Larger $b_i$ corresponds to smaller values of $Q/r_0$, $M/r_0$, and $M/Q$.
We notice that this family of wormholes differs from the one reported in \cite{Blazquez-Salcedo:2020czn}, where the condition $\gamma=1$ was imposed instead of (\ref{Wcond}). The condition (\ref{Wcond}) is more relevant, since the Maxwell equation \arxivonly{(\ref{Maxwell})} has no discontinuity at the throat because of changing the sign of $V''(0)$.

We conclude that the junction conditions for such symmetric wormholes lead to nonsmooth geometries and configuration of matter fields.
Although the metric tensor and matter fields are continuous, their higher derivatives have discontinuity at the throat.
The geometries considered in \cite{Blazquez-Salcedo:2020czn} have additional discontinuity of $V''(0)$. Nevertheless, the Riemann tensor is continuous in both cases.

\section{Smooth asymmetric wormholes}
In order to avoid the above discontinuities, instead of the matching (\ref{matching}) at the throat, we substitute
\begin{equation}\label{smoothfuncs}
F(x) \to \F(x)\equiv\kappa F(x), \quad  B(x) \to \B(x)\equiv\kappa B(x),
\end{equation}
in the Einstein-Dirac-Maxwell equations\arxivonly{ (\ref{finaleqs})}\footnote{Equivalently, we can substitute $G(x)\to\G(x)\equiv\kappa G(x)$.}
and search for a solution in the complete region $-1<x<1$. The equations for the tilted functions do not depend on $\kappa$.
Therefore, we fix $\kappa=1$ in a uniform way\footnote{The physically equivalent solution for $\kappa=-1$ can be obtained by changing the sign of the functions $F(x)$ and $B(x)$ and mirroring the configuration by replacing $x\to-x$.} for the solution in the whole space $-1<x<1$.
The obtained solutions for $N(x)$, $B(x)$, $V(x)$, $F(x)$ and $G(x)$ are smooth functions everywhere, so that the densities of the electromagnetic and Dirac fields do not have discontinuities, and the resulting configuration does not require posing any shell of matter at the throat. The function $B(x)$ changes its sign at the throat ($x=0$) where it crosses the $x$-axis. The latter condition follows from the continuity of the radial vielbein \arxivonly{(see Appendix)} \cite{Cariglia:2018rhw}.

\begin{compactfigure}
\resizebox{\linewidth}{!}{\includegraphics{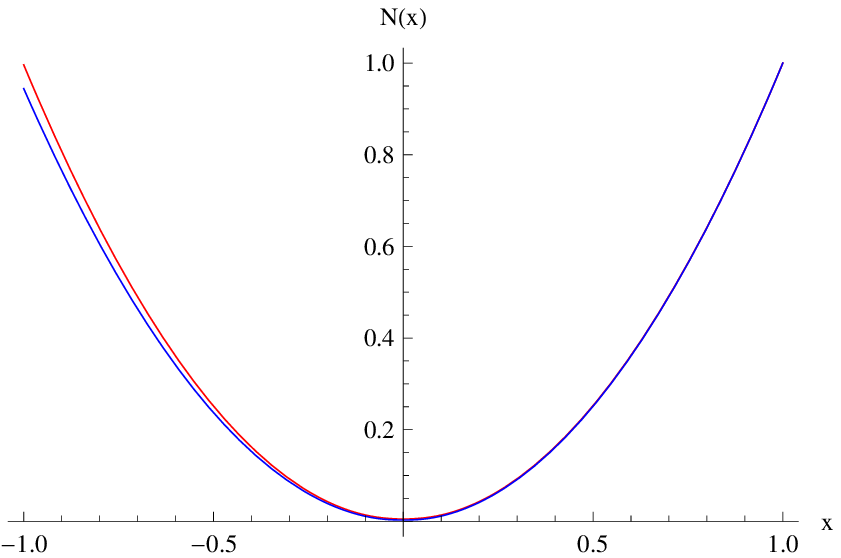}\includegraphics{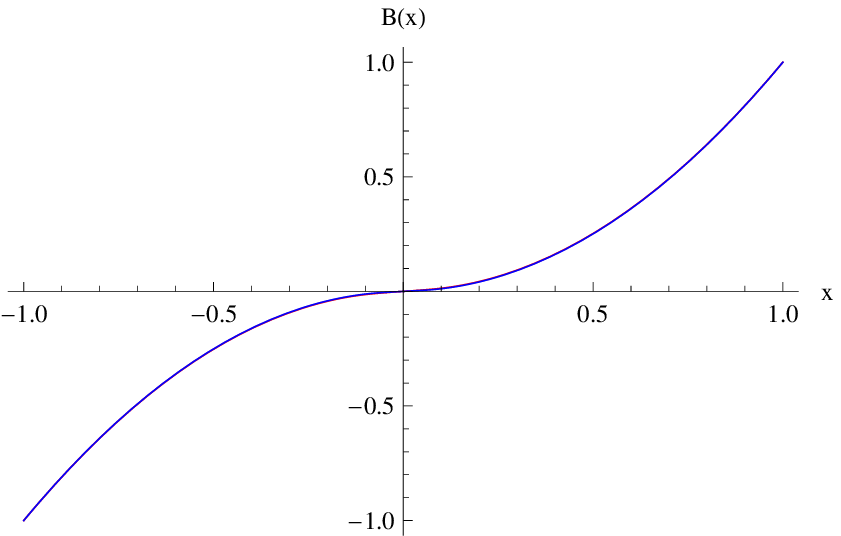}}
\resizebox{\linewidth}{!}{\includegraphics{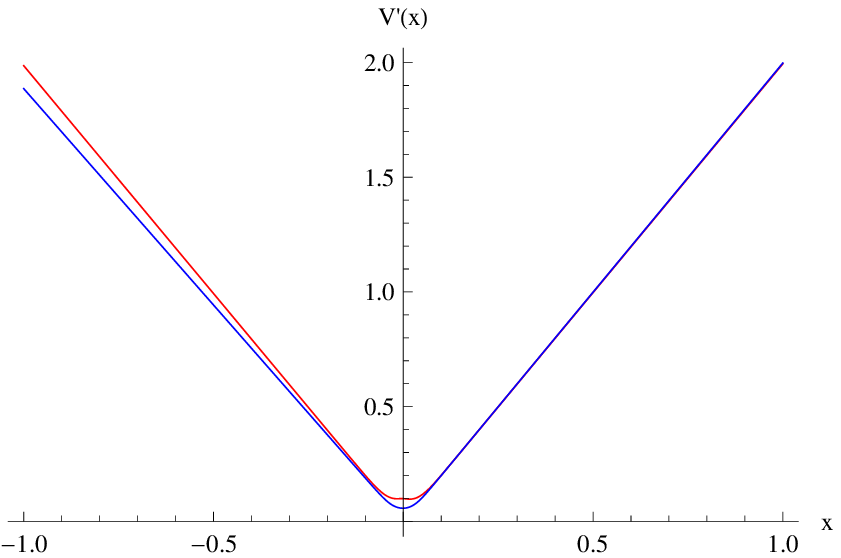}\includegraphics{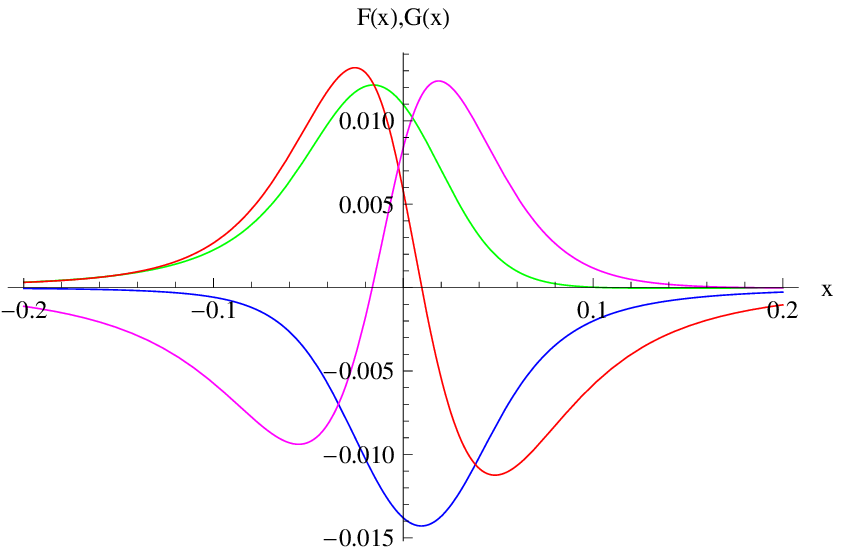}}
\caption{Comparison of two solutions ($qr_0=0.03$, $\mu r_0=0.2$) with the same metric conditions at the throat ($b_i=0.1$) for $N'(0)=0$ and different fermionic configurations: (1) $g_i\approx0.0109$ (green) and $f_i\approx-0.0138$ (blue); (2) $g_i\approx0.0084$ (magenta) and $f_i\approx0.0058$ (red). Even though the metric functions and asymptotic parameters are close for both configurations, the electric and fermionic fields differ significantly near the throat.}\label{fig:simcond}
\end{compactfigure}
Since we do not have $Z_2$ symmetry with respect to the throat, we need to impose two asymptotic conditions independently. For fixed values of $n_i$ and $b_i$ we find that, depending on the choice of $f_i$ and $g_i$, $F(x)$ and $G(x)$ can have the following asymptotic behavior \arxivonly{(see Fig.~\ref{fig:b01initialregion})}:
\begin{enumerate}
\item $\displaystyle\lim_{x\to\pm1}G(x)=+\infty, \lim_{x\to\pm1}F(x)=\mp\infty$ (red); \\
\item $\displaystyle\lim_{x\to\pm1}G(x)=-\infty, \lim_{x\to\pm1}F(x)=\pm\infty$ (blue); \\
\item $\displaystyle\lim_{x\to\pm1}G(x)=\pm\infty, \lim_{x\to\pm1}F(x)=-\infty$ (magenta); \\
\item $\displaystyle\lim_{x\to\pm1}G(x)=\mp\infty, \lim_{x\to\pm1}F(x)=+\infty$ (green). \\
\end{enumerate}
When $F(x)$ and $G(x)$ change the sign there are exponentially decaying solutions, which we find by shooting $f_i$ and $g_i$.
The resulting metric is asymptotically flat on both sides of the throat,
\begin{eqnarray}
\nonumber
N(x)&=&\sigma_{\pm}\left(1-\dfrac{2M_{\pm}}{r_0}(1\mp x)+\Order{1\mp x}^2\right),\\
\label{asymasymptotic}
B(x)&=&\pm\left(1-\gamma_{\pm}\dfrac{2M_{\pm}}{r_0}(1\mp x)+\Order{1 \mp x}^2\right),\\
\nonumber
V'(x)&=&\sigma_{\pm}\left(\pm\dfrac{2Q_{\pm}}{r_0}+\Order{1\mp x}\right).
\end{eqnarray}

Note that, due to asymmetry, $\sigma_+\neq\sigma_-$, two stationary asymptotic observers on the opposite sides of the wormhole's throat have relativistic time dilation (redshift). By scaling $n_i$ we can fix the coordinate time according to one of the observers, so that either $\sigma_+=1$ or $\sigma_-=1$. In addition, unless $q=0$, we have $M_+\neq M_-$ and $Q_+\neq Q_-$. However, for all the solutions we have obtained, $\gamma_+=\gamma_-\approx1$, at least within the numerical accuracy.

Since the electric potential $V(x)$ is now a smooth function everywhere, requirement $V''(0)=0$ (\ref{Wcond}) seems not relevant for this case. Although we have obtained the wormholes, which have both smooth asymmetric continuation and a continuous symmetric one, the most physically relevant condition on the throat is, apparently (cf.~\cite{Bolokhov:2021fil}),
\begin{equation}\label{Acond}
N'(0)=0,
\end{equation}
which leads to no gravitational force experienced by a stationary observer at the throat ($x=0$). The corresponding time delation is given by $\sigma_0=N(0)=n_i$.

Since the field equations \arxivonly{(\ref{finaleqs})} are invariant under changing sign of the fermionic functions $G(x)\to-G(x)$ and $F(x)\to-F(x)$, we can study only $g_i>0$ without loss of generality. For a given $b_i$ there are various possible solutions, corresponding to different values $f_i$ and $g_i$. We compare two such solutions on Fig.~\ref{fig:simcond}. The largest absolute values of $g_i$ and $f_i$ (producing asymptotically flat solutions) with opposite signs (solution (1) in Fig.~\ref{fig:simcond}) correspond to the fermion configuration with $G(x)\neq0$ and $F(x)\neq0$. The second largest values of $g_i$ and $f_i$ lead to the fermion configuration, for which both $G(x)$ and $F(x)$ cross zero once near the throat. The closer the solution to the origin, the more zeroth the fermionic functions possess and, consequently, the integration and shooting should be performed with higher accuracy.

For fixed fermion mass $\mu$ and charge $q$ we obtain different wormhole solutions by varying $b_i$. In the limit $b_i\to0$ we approach the extreme Reissner-Nordström black hole. The asymptotic charge of the wormholes is always larger than the asymptotic mass for all the obtained solutions: $Q_+>M_+$ and $Q_->M_-$. Unfortunately, considerable increasing of $b_i$ makes integration less stable and requires further increasing of precision for the whole procedure.

\begin{compactfigure}
\resizebox{\linewidth}{!}{\includegraphics{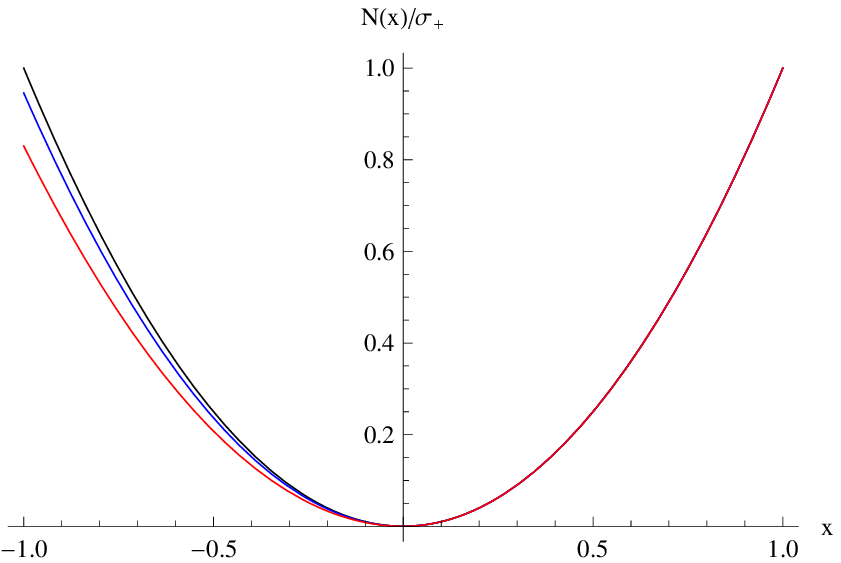}\includegraphics{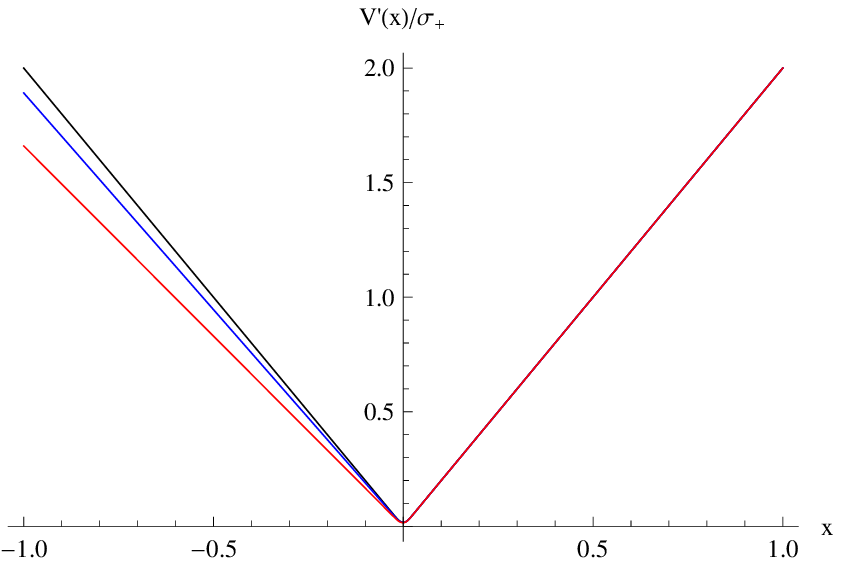}}
\caption{Comparison of solutions of the same fermion mass ($\mu r_0=0.2$) for $b_i=0.03$: $q=0$, $g_i\approx0.003346$, $f_i\approx-0.004082$ (black, top); $qr_0=0.03$, $g_i\approx0.003286$, $f_i\approx-0.004131$ (blue); $qr_0=0.1$, $g_i\approx0.003145$, $f_i\approx-0.004247$ (red, bottom). While $G(x)$, $F(x)$, and $B(x)$ are very close, the lapse function $N(x)$ and the electric field $V'(x)$ differ significantly on one of the asymptotics.}\label{fig:unchargedcomp}
\end{compactfigure}

We also obtained the wormhole solutions supported by uncharged fermions ($qr_0=0$, $\mu r_0=0.2$). Since the Compton wavelength for the neutrinos can be of the order of millimeters \cite{Choudhury:2018byy}, such solutions include traversable wormholes of the millimeter throat size. We have found that the asymptotic mass and charge are the same for both sides of the throat. However, the geometry is asymmetric and there is a small redshift between the asymptotic observers $\sigma_-/\sigma_+\approx0.99993$. We compare these solutions on Fig.~\ref{fig:unchargedcomp}. As the wormhole geometry under consideration is quite different from that of the Schwarzschild black hole, electromagnetic and gravitational radiation in the vicinity of such wormholes could be potentially observable and distinctive from those for the Schwarzschild case. The first step in the direction has been made in \cite{Churilova:2021tgn}, where the shadows, quasinormal ringing, and echoes of the wormholes were considered. However, more realistic models must also include wormholes' rotation.

\section{Conclusions}
In \cite{Blazquez-Salcedo:2020czn} the first classical configuration supporting asymptotically flat traversable and symmetric relative the throat wormholes in the Einstein theory has been found in the presence of Maxwell and two Dirac fields with usual (nonexotic) coupling. Nevertheless, those classical wormhole configurations have a number of unfeasible physical properties of matter at the wormhole throat, such as coexistence of particles and antiparticles without annihilation, nonsmoothness of matter fields, etc. Here we found solutions describing asymmetric asymptotically flat traversable wormholes supported by smooth metric and matter fields, which, therefore, are free of all the above problems. This gives us the hope that such kind of wormholes could exist in nature.

\begin{acknowledgments}
We thank J.~Blázquez-Salcedo for sharing his numerical data, K.~Bronnikov for bringing our attention to \cite{Bolokhov:2021fil}, and R.~Wald and R.~Weinbaum for useful discussions. We acknowledge support of the Grant No.~19-03950S of Czech Science Foundation (GAČR).
\end{acknowledgments}

\arxivonly{
\appendix\section{Supplemental Material}
\begin{table*}
\begin{tabular}{|c|c|c|c|c|c|c|c|c|c|}
\hline
$b_i$ & $g_i$ & $f_i$ & $n_i/\sigma_+$ & $\sigma_-/\sigma_+$ & $Q_+/r_0$ & $Q_-/r_0$ & $M_+/r_0$ & $M_-/r_0$ & $\omega r_0/\sigma_+$ 
\\
\hline
$0.03$ & $0.003286$ & $-0.004131$ & $0.000257$ & $0.945685$ & $0.999894$ & $0.999883$ & $0.999894$ & $0.999883$ & $-0.0003901$ 
\\
$0.04$ & $0.004382$ & $-0.005509$ & $0.000457$ & $0.945577$ & $0.999811$ & $0.999792$ & $0.999811$ & $0.999792$ & $-0.0006936$ 
\\
$0.05$ & $0.005478$ & $-0.006888$ & $0.000714$ & $0.945436$ & $0.999705$ & $0.999676$ & $0.999705$ & $0.999676$ & $-0.0010838$ 
\\
$0.06$ & $0.006575$ & $-0.008268$ & $0.001028$ & $0.945261$ & $0.999575$ & $0.999533$ & $0.999575$ & $0.999533$ & $-0.0015608$ 
\\
$0.07$ & $0.007672$ & $-0.009649$ & $0.001400$ & $0.945053$ & $0.999422$ & $0.999365$ & $0.999421$ & $0.999364$ & $-0.0021247$ 
\\
$0.08$ & $0.008770$ & $-0.011032$ & $0.001830$ & $0.944816$ & $0.999246$ & $0.999171$ & $0.999244$ & $0.999169$ & $-0.0027754$ 
\\
\hline
$0.03$ & $0.002498$ & $0.0017518$ & $0.000442$ & $0.995020$ & $0.999743$ & $0.999732$ & $0.999743$ & $0.999732$ & $-0.0011131$ 
\\
$0.04$ & $0.003332$ & $0.0023351$ & $0.000786$ & $0.995076$ & $0.999543$ & $0.999524$ & $0.999542$ & $0.999523$ & $-0.0019789$ 
\\
$0.05$ & $0.004167$ & $0.0029179$ & $0.001230$ & $0.995146$ & $0.999285$ & $0.999256$ & $0.999284$ & $0.999255$ & $-0.0030926$ 
\\
$0.06$ & $0.005004$ & $0.0035001$ & $0.001773$ & $0.995230$ & $0.998970$ & $0.998928$ & $0.998969$ & $0.998926$ & $-0.0044544$ 
\\
$0.07$ & $0.005844$ & $0.0040815$ & $0.002416$ & $0.995328$ & $0.998598$ & $0.998540$ & $0.998595$ & $0.998537$ & $-0.0060646$ 
\\
$0.08$ & $0.006685$ & $0.0046619$ & $0.003160$ & $0.995444$ & $0.998169$ & $0.998092$ & $0.998163$ & $0.998087$ & $-0.0079238$ 
\\
\hline
\end{tabular}
\caption{Solution parameters ($qr_0=0.03$, $\mu r_0=0.2$) for the families of wormholes.}\label{tabl:solutionpars}
\end{table*}

The curved-space gamma matrices $\gg^{\mu}$ are defined as
\begin{equation}\label{gg}
\gg_{\mu}\equiv e_{\mu\alpha}\ge^{\alpha},
\end{equation}
where $e_{\mu\alpha}$ is a tetrad of vectors,
$$e_{\mu\alpha}e_{\nu\beta}\eta^{\alpha\beta}=g_{\mu\nu},$$
and $\ge^{\alpha}$ are the Dirac matrices, satisfying
\begin{equation}\label{Diracmatrices}
\ge^{\alpha}\ge^{\beta}+\ge^{\beta}\ge^{\alpha}=2\eta^{\alpha\beta}I.
\end{equation}

\begin{figure}
\resizebox{\linewidth}{!}{\includegraphics{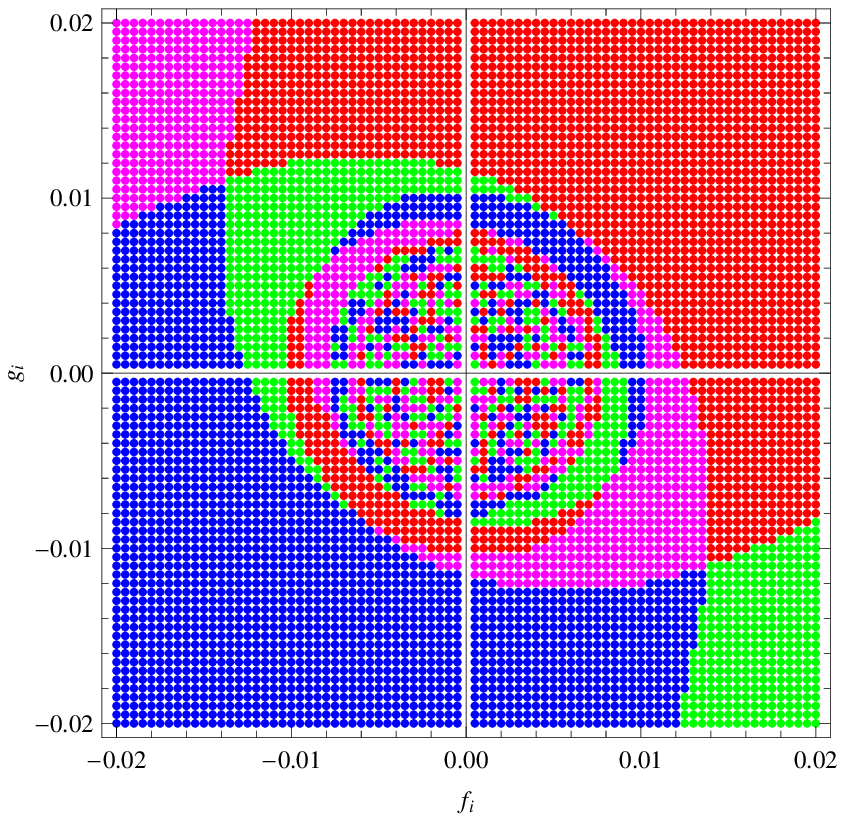}}
\caption{Asymptotic behavior of the fermions ($qr_0=0.03$, $\mu r_0=0.2$) for the asymmetric wormhole solutions ($N'(0)=0$) for $b_i=0.1$. All the points where the above four regions meet, i.~e. both $F(x)$ and $G(x)$ change the sign at both spatial infinities, correspond to the asymptotically flat solutions for which both $G(\pm1)=0$ and $F(\pm1)=0$. For small values of $f_i$ and $g_i$, a fine structure of the regions is needed in order to determine such points.}\label{fig:b01initialregion}
\end{figure}

The spinor covariant derivative is
\begin{equation}\label{Diracadjoint}
\D_{\mu}\equiv\partial_{\mu} - \Gamma_\mu - \imo qA_{\mu},
\end{equation}
where $q$ is the Dirac field charge and spinor connection matrices $\Gamma_\mu$,
\begin{equation}\label{spinorconnection}
\Gamma_\mu=-\frac{1}{4}\omega_{\mu\alpha\beta}\ge^{\alpha}\ge^{\beta},
\end{equation}
are calculated making use of the spin connection
\begin{equation}\label{spinconnection}
\omega_{\mu\alpha\beta}=(e_{\nu\alpha}\Gamma^{\nu}_{\mu\rho}-\partial_{\mu}e_{\rho\alpha})g^{\rho\lambda}e_{\lambda\beta},
\end{equation}
where $\Gamma^{\nu}_{\mu\rho}$ is the Levi-Civita connection,
\begin{equation}\label{Levi-Civita}
\Gamma^{\nu}_{\mu\rho}=\frac{1}{2}g^{\nu\lambda}(\partial_{\mu}g_{\rho\lambda}+\partial_{\rho}g_{\mu\lambda}-\partial_{\lambda}g_{\mu\rho}).
\end{equation}

By varying the action (\ref{action}), we find the following equations:
\begin{eqnarray}\label{Dirac}
-\frac{4\pi\imo}{\sqrt{-g}}\frac{\delta S}{\delta\PD_{\epsilon}}&=&\left(\gg^{\nu}\D_{\nu}-\mu\right)\P_{\epsilon}=0;\\
\label{Maxwell}
\frac{4\pi}{\sqrt{-g}}\frac{\delta S}{\delta A_{\mu}}&=&\frac{\partial_{\nu}\sqrt{-g}F^{\mu\nu}}{2\sqrt{-g}}-q j^{\mu}=0,\\
\label{current}
j^{\mu}&\equiv&\PD_1\gg^{\mu}\P_1+\PD_2\gg^{\mu}\P_2;\\
\label{Einstein}
\frac{8\pi}{\sqrt{-g}}\frac{\delta S}{\delta g^{\mu\nu}}&=&\frac{1}{2}\left(R_{\mu\nu}-\frac{1}{2}g_{\mu\nu}R\right)-T_{\mu\nu}=0,
\end{eqnarray}
where the stress-energy tensor is
$$T_{\mu\nu}=T^M_{\mu\nu}+T^1_{\mu\nu}+T^2_{\mu\nu},$$
with the Maxwell and Dirac-field stress-energy tensors defined as follows:
\begin{eqnarray}\label{DiracM}
T^M_{\mu\nu}&=&F_{\mu\lambda}F_{\nu\rho}g^{\lambda\rho}-\frac{1}{4}g_{\mu\nu}F_{\lambda\rho}F^{\lambda\rho},\\
\label{DiracT}
T^{\epsilon}_{\mu\nu}&=&\im{\PD_{\epsilon}\gg_{\mu}\D_{\nu}\P_{\epsilon}+\PD_{\epsilon}\gg_{\nu}\D_{\mu}\P_{\epsilon}}.
\end{eqnarray}

Following \cite{Dolan:2015eua}, we choose representation for the Dirac matrices,
\begin{equation}\label{representation}
\begin{array}{ll}
  \ge^0=\imo\left(
              \begin{array}{cc}
                0 & I \\
                I & 0 \\
              \end{array}
            \right),
&
  \ge^1=\imo\left(
              \begin{array}{cc}
                0 & \sigma_3 \\
                -\sigma_3 & 0 \\
              \end{array}
            \right),
\\&\\
  \ge^2=\imo\left(
              \begin{array}{cc}
                0 & \sigma_1 \\
                -\sigma_1 & 0 \\
              \end{array}
            \right),
 &
  \ge^3=\imo\left(
              \begin{array}{cc}
                0 & \sigma_2 \\
                -\sigma_2 & 0 \\
              \end{array}
            \right),
 \end{array}
 \end{equation}
 where $\sigma_i$ denote the standard Pauli matrices,
 $$
 \sigma_1=\left(
              \begin{array}{cc}
                0 & 1 \\
                1 & 0 \\
              \end{array}
            \right),
 \quad
 \sigma_2=\left(
              \begin{array}{cc}
                0 & -\imo \\
                \imo & 0 \\
              \end{array}
            \right),
 \quad
 \sigma_3=\left(
              \begin{array}{cc}
                1 & 0 \\
                0 & -1 \\
              \end{array}
            \right).
 $$
so that
\begin{equation}\label{gamma5}
  \ge^5=\imo\ge^0\ge^1\ge^2\ge^3=\left(
              \begin{array}{cc}
                -I & 0 \\
                0 & I \\
              \end{array}
            \right),
\end{equation}
and the projection operator $P_{\pm}=\frac{1}{2}(I\pm\ge^5)$ implies that any Dirac four-spinor is
\begin{equation}\label{DiracSpinor}
\P=\left(\begin{array}{c}\psi_-\\\psi_+\end{array}\right),
\end{equation}
where $\psi_+$ and $\psi_-$ are right- and left-handed two-spinors, respectively.

Its Dirac adjoint is defined as
\begin{equation}\label{DiracAdjoint}
\PD\equiv-\P^\dagger\ge^0=-\imo\left(\begin{array}{cc}\psi_+^\dagger & \psi_-^\dagger\end{array}\right),
\end{equation}
so that
$$\imo\PD\P=\psi_+^\dagger\psi_-+\psi_-^\dagger\psi_+.$$

We choose the tetrads
\begin{equation}\label{vielbein}
e_{\mu\alpha}=\left(
                  \begin{array}{cccc}
                    -N(x) & 0 & 0 & 0 \\
                    0 & r'(x)/B(x) & 0 & 0 \\
                    0 & 0 & r(x) & 0 \\
                    0 & 0 & 0 & r(x)\sin\theta \\
                  \end{array}
                \right).
\end{equation}
Notice that as $N(x)$, $B(x)$ and $r(x)$ are smooth across the throat $x=0$ for the asymmetric wormhole solution, the tetrads are smooth as well.

Then, after some algebra, equations (\ref{Dirac}), (\ref{Maxwell}), and (\ref{Einstein}) are reduced to the following set of differential equations:
\begin{widetext}
\begin{subequations}\label{finaleqs}
\begin{eqnarray}
F'(x)&=&\kappa\frac{r'(x)}{B(x)}\left(\frac{8 F(x)^2 G(x)}{ B(x)}-\frac{F(x)}{r(x)}\right)-\frac{F(x)r'(x)}{4 B(x)^2 r(x)}-\frac{3 F(x) r'(x)}{4 r(x)}+\frac{F(x) V'(x)^2 r(x)}{4 N(x)^2 r'(x)}
\\\nonumber&&
+\frac{\omega+qV(x)}{N(x)}\frac{r'(x)}{B(x)}\left(G(x)-\frac{4 F(x) r(x) \left(F(x)^2+G(x)^2\right)}{B(x)}\right)-\mu\frac{r'(x)}{B(x)}\left(G(x)+\frac{4 F(x) r(x)\left(F(x)^2-G(x)^2\right)}{B(x)}\right),\\
G'(x)&=&\kappa \frac{r'(x)}{B(x)}\left(\frac{8 F(x) G(x)^2}{B(x)}+\frac{G(x)}{r(x)}\right)-\frac{G(x)r'(x)}{4 B(x)^2 r(x)}-\frac{3 G(x) r'(x)}{4 r(x)}+\frac{G(x) V'(x)^2 r(x)}{4 N(x)^2 r'(x)}
\\\nonumber&&
-\frac{\omega+qV(x)}{N(x)}\frac{r'(x)}{B(x)}\left(F(x)+\frac{4 G(x) r(x) \left(F(x)^2+G(x)^2\right)}{B(x)}\right)-\mu\frac{r'(x)}{B(x)}\left(F(x)+\frac{4 G(x) r(x) \left(F(x)^2-G(x)^2\right)}{B(x)}\right),\\
V''(x)&=&V'(x)\left(\frac{r''(x)}{r'(x)}-\frac{2 r'(x)}{r(x)}-\kappa\frac{16 F(x) G(x) r'(x)}{ B(x)^2}\right)+\frac{8q N(x)r'(x)^2}{B(x)^2}\left(F(x)^2+G(x)^2\right)
\\\nonumber&&
+\frac{16V'(x)r'(x)r(x)}{B(x)^2}\left(\frac{\omega+qV(x)}{N(x)}\left(F(x)^2+G(x)^2\right)+\frac{\mu}{2} \left(F(x)^2 -G(x)^2\right)\right),\\
N'(x)&=&N(x)r'(x)\left(\frac{1-B(x)^2}{2 B(x)^2 r(x)}-\kappa\frac{16 F(x) G(x)}{B(x)^2}\right)-\frac{V'(x)^2 r(x)}{2 N(x) r'(x)}
\\\nonumber&&
+\frac{8N(x)r'(x)r(x)}{B(x)^2}\left(\frac{\omega+qV(x)}{N(x)}\left(F(x)^2+G(x)^2\right)+\mu\left(F(x)^2-G(x)^2\right)\right),\\
B'(x)&=&r'(x)\frac{1-B(x)^2}{2 B(x) r(x)}-\frac{B(x) V'(x)^2 r(x)}{2 N(x)^2 r'(x)}-\frac{8r(x)r'(x)}{B(x)}\frac{\omega+qV(x)}{N(x)}\left(F(x)^2+G(x)^2\right).
\end{eqnarray}
\end{subequations}
\end{widetext}


It is convenient to introduce the auxiliary functions
\begin{eqnarray}
\label{U(x)}
U(x)&\equiv&\omega+qV(x),
\\\label{W(x)}
W(x)&\equiv&V'(x),
\end{eqnarray}
therefore
\begin{equation}\label{Uequation}
U'(x)=qW(x),
\end{equation}
and (\ref{finaleqs}) is reduced to a system of ordinary differential equations of the first order.

We introduce the compact coordinate through
\begin{equation}\label{r(x)}
r(x)=\frac{r_0}{1-x^2},
\end{equation}
and assume that all the functions can be expanded in series at the throat ($x=0$).

Since $r'(0)=0$, we require that $B(0)=0$. When expanding the righthand sides of (\ref{finaleqs}) in series at the throat we obtain divergent terms, which must vanish for the regular solutions. These regularity conditions at the throat lead to the following relations:
\begin{eqnarray}\label{W0}
W(0)^2&=&\frac{2N(0)^2}{B'(0)^2}\Biggl(2-B'(0)^2
\\\nonumber&&+16r_0^2\mu\left(F(0)^2-G(0)^2\right)-32r_0\kappa F(0)G(0)\Biggr),
\\
\label{U0}
U(0)&=&-\frac{N(0)}{32r_0^2\left(F(0)^2+G(0)^2\right)}\Biggl(B'(0)^2
\\\nonumber&&+16r_0^2\mu\left(F(0)^2-G(0)^2\right)-32r_0\kappa F(0)G(0)\Biggr).
\end{eqnarray}
The choice for the sign of $W(0)$ corresponds to the sign of the asymptotic charge. Without loss of generality we assume that $W(0)>0$. We also impose $V(0)=0$ so that $U(0)=\omega$.

Thus, behavior of the solution at the throat ($x=0$) is completely determined by the parameters $n_i$, $b_i$, $f_i$ and $g_i$ introduced in (\ref{initialcond}). The value of $n_i$ scales the units of time and can be chosen in such a way that the coordinate $t$ is the time according to one of the asymptotic observers. For given $b_i$ we find several asymptotically flat asymmetric wormhole solutions, corresponding to different values $f_i$ and $g_i$ (see Fig.~\ref{fig:b01initialregion}).

The values of parameters describing the wormhole configurations are given in table~\ref{tabl:solutionpars}. Notice that, in \cite{Blazquez-Salcedo:2020czn} the family of wormhole solutions was parametrized by $\omega$, while we parametrize the family by $b_i$, which does not depend on the choice of the asymptotic time. We see that $\omega/\sigma_+$ depends monotonically on $b_i$.

}

\end{document}